\documentclass[english,twocolumn]{revtex4}
\usepackage[T1]{fontenc}
\usepackage[latin9]{inputenc}
\usepackage{verbatim}
\usepackage{amsbsy}
\usepackage{amstext}
\usepackage{graphicx}
\usepackage{esint}

\makeatletter

\@ifundefined{textcolor}{}
{%
 \definecolor{BLACK}{gray}{0}
 \definecolor{WHITE}{gray}{1}
 \definecolor{RED}{rgb}{1,0,0}
 \definecolor{GREEN}{rgb}{0,1,0}
 \definecolor{BLUE}{rgb}{0,0,1}
 \definecolor{CYAN}{cmyk}{1,0,0,0}
 \definecolor{MAGENTA}{cmyk}{0,1,0,0}
 \definecolor{YELLOW}{cmyk}{0,0,1,0}
}

\usepackage[percent]{overpic}

\makeatother

\usepackage{babel}
\begin{document}
\global\long\def\braketop#1#2#3{\left\langle #1\vphantom{#2}\vphantom{#3}\left|#2\vphantom{#1}\vphantom{#3}\right|#3\vphantom{#1}\vphantom{#2}\right\rangle }
\global\long\def\braket#1#2{\left\langle \left.#1\vphantom{#2}\right|#2\right\rangle }
\global\long\def\bra#1{\left\langle #1\right|}
\global\long\def\ket#1{\left|#1\right\rangle }
\global\long\def\ketbra#1#2{\left|#1\vphantom{#2}\right\rangle \left\langle \vphantom{#1}#2\right|}
\global\long\def\valley{K^{\prime}}
\global\long\def\othervalley{K}
\global\long\def\figlab#1{(#1)}
\global\long\def\vec#1{\mathbf{#1}}

\title{Husimi Maps in Graphene}

\author{Douglas J. Mason, Mario F. Borunda, and Eric J. Heller}

\affiliation{Department of Physics, Harvard University, Cambridge, MA 02138, USA}
\begin{abstract}
We present a method for bridging the gap between the Dirac effective
field theory and atomistic simulations in graphene based on the Husimi
projection, allowing us to depict phenomena in graphene at arbitrary
scales. This technique takes the atomistic wavefunction as an input,
and produces semiclassical pictures of quasiparticles in the two Dirac
valleys. We use the Husimi technique to produce maps of the scattering
behavior of boundaries, giving insight into the properties of wavefunctions
at energies both close to and far from the Dirac point. Boundary conditions
play a significant role to the rise of Fano resonances, which we examine
using the Husimi map to deepen our understanding of bond currents
near resonance.
\end{abstract}
\maketitle

\section{Introduction}

With interest and experimental capabilities in graphene devices
growing\citep{Geim-Nature,Experiment1,Experiment2,Experiment3,Experiment5,Experiment4,Mario,Mario2-not-Mario},
the need has never been greater to improve our understanding of quantum
states in this material. %
Despite the success of the Dirac effective field theory for graphene\citep{castro-neto},
however, many technological proposals arise from predictions using
the more fundamental tight-binding approximation\citep{Wimmer-Spin-Currents-Rough-Nanoribbons,KaxirasNanoFlakes,Wimmer-Spin-Transport,WeiLi}.
This is because the atomistic model that underlies the Dirac theory
is able to incorporate phenomena such as scattering from small defects\citep{molecular-doping-graphene,imaging-edge-states,defect-graphene,defect-point-standing-wave,defects-graphene-point-theory-predict},
ripples\citep{katsnelson-ripples}, or edge types\citep{reconstructing-edges,imaging-edges,edge-types-effects}
-- all of which promise technological applications. However, atomistic
calculations are computationally expensive, and replacing these features
with scattering theories in a more-efficient Dirac model introduces
substantial challenges. A robust approach which can analyze the atomistic
wavefunction to produce semiclassical pictures of quasiparticles in
the two Dirac valleys remains to be seen. %

To address these issues and expand our understanding of graphene quantum
states, we use the Husimi projection technique, introduced by Mason
\emph{et al.} \citep{Mason-PRL,Mason-Husimi-Continuous,Mason-Husimi-Lattices},
to produce snapshots of the local momentum distribution and underlying
semiclassical structure in graphene wavefunctions. When Husimi projections
are calculated at many points across a system, the Husimi map that
results provides a semiclassical picture of the atomistic wavefunction.
In this article, we define the Husimi map for graphene systems (Sec.~\ref{sec:Method}),
and use it to deepen our understanding of boundary conditions in both
high-energy relativistic scar states\citep{PhysRevLett.53.1515,Huang-PRL-Scars}
(Sec.~\ref{sub:Realtivistic-Scars}), and states near the Dirac point
(Sec.~\ref{sub:Low-Energy-States}). We then use Husimi maps to interpret
Fano resonances\citep{fano-original,JJAP.36.3944,fano-review} within
this novel material (Sec.~\ref{sub:Fano-Resonance}).

\section{Method\label{sec:Method}}

\subsection{Definition of the Husimi Projection\label{sub:Definition-of-the}}

The conduction band of the graphene system can be approximated as
a honeycomb lattice with a single $p_{z}$ orbital located at each
carbon-atom lattice site\citep{castro-neto}. The Husimi function
is defined as a measurement between a wavefunction $\psi(\left\{ \vec r_{i}\right\} )$
defined at each orbital, and a coherent state $\ket{\vec r_{0},\vec k_{0},\sigma}$
which describes an envelope function over those sites that minimizes
the joint uncertainty in spatial and momentum coordinates. The parameter
$\sigma$ defines the spatial spread of the coherent state and defines
the uncertainties in space and momentum according to the well-known
relation
\begin{equation}
\Delta x\propto\frac{1}{\Delta k}\propto\sigma.
\end{equation}
As a result, there is a trade-off for any value of $\sigma$ selected:
for small $\sigma$, there is better spatial resolution but poorer
resolution in $k$-space, and vice versa for large $\sigma$. 

Writing out the dot product of the wavefunction and the coherent state

\begin{eqnarray}
\braket{\psi}{\vec r_{0},\vec k_{0},\sigma} & = & \left(\frac{1}{\sigma\sqrt{\pi/2}}\right)^{\text{}}\nonumber \\
 &  & \times\sum_{i}\psi\left(\vec r_{i}\right)e^{-\left(\vec r_{i}-\vec r_{0}\right)^{2}/4\sigma^{2}+i\vec k_{0}\cdot\vec r_{i}},\label{eq:Gaussian-Function}
\end{eqnarray}
the Husimi function is defined as
\begin{equation}
\text{Hu}\left(\vec r_{0},\vec k_{0},\sigma;\psi(\left\{ \vec r_{i}\right\} )\right)=\left|\braket{\psi}{\vec r_{0},\vec k_{0},\sigma}\right|^{2}.\label{eq:Husimi}
\end{equation}

Weighting the Husimi function by the wavevector $\vec k_{0}$ produces
the $k$-space Husimi vector, and weighting it by the group velocity
vector $\boldsymbol{\nabla}_{\vec k}E\left(\vec k'\right)$ produces
the group-velocity Husimi vector. The latter is a stronger reflection
of classical dynamics in the system, and is used for all results in
this paper. At each point in the system, we can sweep through $k$-space
by rotating the wavevector $\vec k_{0}$ along the Fermi surface in
the dispersion relation. The multiple Husimi vectors which result
form the full Husimi projection, providing a snapshot of the local
momentum distribution. This paper uses 32 wavevectors along the Fermi
surface of two-dimensional graphene to produce group-velocity Husimi
projections\citep{Mason-Husimi-Lattices}.

Even though a few plane waves may dominate the wavefunction, momentum
uncertainty of the coherent state can result in many non-vanishing
Husimi vectors. Assuming that the dominant plane waves at a point
are sufficiently separated in $k$-space, it is possible to recover
their wavevectors using the Multi-Modal Algorithm in Mason \emph{et
al.}\citep{Mason-Husimi-Lattices}. This method singles out the important
wavevectors contributing to a wavefunction at each point; in this
paper, we additionally remove results below a certain threshold to
clarify our results.

The integral over Husimi vectors at a single point defines a new vector-valued
function $\vec{Hu}\left(\vec r_{0},\sigma;\psi(\left\{ \vec r\right\} )\right)$,
which is equal to
\begin{equation}
\vec{Hu}\left(\vec r_{0},\sigma;\psi\left(\left\{ \vec r_{i}\right\} \right)\right)=\int\left|\braket{\psi}{\vec r_{0},\vec k_{0},\sigma}\right|^{2}\vec k_{0}d^{d}k_{0}.\label{eq:Husimi-Vector-Old}
\end{equation}
It has been shown that for $\sigma k\ll1$, this function is equal
to the flux operator\citep{Mason-Husimi-Continuous}. To better represent
the classical dynamics of the system we can instead weight the integrand
by the group velocity $\boldsymbol{\nabla}_{\vec k}E\left(\vec k'\right)$
to obtain the group-velocity Husimi flux $\vec{Hu}_{\text{g}}\left(\vec r_{0},\sigma;\psi(\vec r)\right)$
equal to
\begin{equation}
\vec{Hu}_{\text{g}}\left(\vec r_{0},\sigma;\psi\left(\left\{ \vec r_{i}\right\} \right)\right)=\int\left|\braket{\psi}{\vec r_{0},\vec k_{0},\sigma}\right|^{2}\boldsymbol{\nabla}_{\vec k}E\left(\vec k_{0}\right)d^{d}k_{0}.\label{eq:Husimi-Vector}
\end{equation}
which is used throughout the paper. 

Even though the Husimi projection is related to the flux operator,
it provides much more information since it can be used on stationary
states which exhibit zero flux, and because it can isolate individual
bands and valleys in the dispersion relation.%
{} Sampling the Husimi projection at many points across a system to
produce a Husimi map, we can produce a much better picture of the
classical dynamics underlying the wavefunction.

\subsection{The Honeycomb Band Structure\label{sub:The-Graphene-Band}}

This paper examines the honeycomb lattice Hamiltonian using the nearest-neighbor
tight-binding approximation 
\begin{equation}
H=\sum_{i}\epsilon_{i}\mathbf{a}_{i}^{\dagger}\mathbf{a}_{i}-t\sum_{\left\langle ij\right\rangle }\mathbf{a}_{i}^{\dagger}\mathbf{a}_{j},
\end{equation}
where $\mathbf{a}_{i}^{\dagger}$ is the creation operator at orbital
site $i$, and we sum over the set of nearest neighbors. To compare
against experiment, the hopping integral value is given by $t=2.7\text{eV}$,
while $\epsilon$ is set to the value of the Fermi energy\citep{Geim-Nature,castro-neto}.
Eigenstates of closed stadium billiard systems are computed using
sparse matrix eigensolvers to produce individual wavefunctions. %

\begin{figure}
\begin{centering}
\includegraphics[width=0.85\columnwidth]{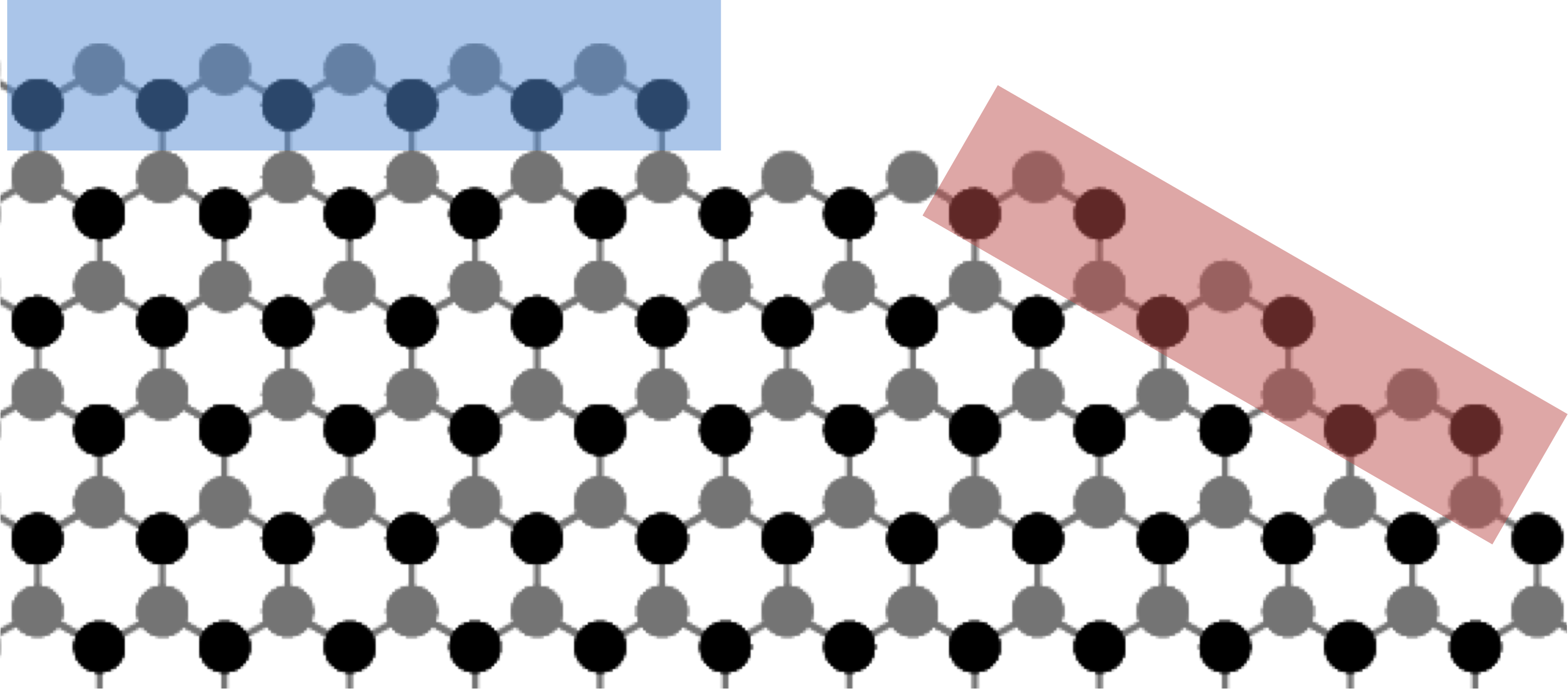}
\par\end{centering}

\caption{\label{fig:Edge-Close-Up}A magnified view of a boundary on a graphene
flake. The orientation of the cut relative to the orientation of the
lattice can produce two edge types, zigzag (highlighted in blue) and
armchair (highlighted in red). The two sublattices of the unit cell
are indicated in black (A-sublattice) and grey (B-sublattice).}
\end{figure}
We study finite graphene systems extracted from an infinite honeycomb
lattice. A filter is applied to remove atom sites which are attached
to only one other atom site, and to bridge under-coordinated sites
whose $\pi$ orbitals would strongly overlap. As a result, each edge
is either a pure zig-zag, armchair, or mixed boundary, as shown in
Fig.~\ref{fig:Edge-Close-Up}. Recent studies have suggested that
under certain circumstances, zig-zag edges reconstruct to form a 5-7
chain\citep{Re-ZagDFT}, however their scattering properties appear
to be identical to regular zig-zag boundaries \citep{Re-ZagWimmer}.
We have elected not to incorporate these features and leave them to
future work. %

\begin{figure}
\begin{centering}
\includegraphics[width=0.85\columnwidth]{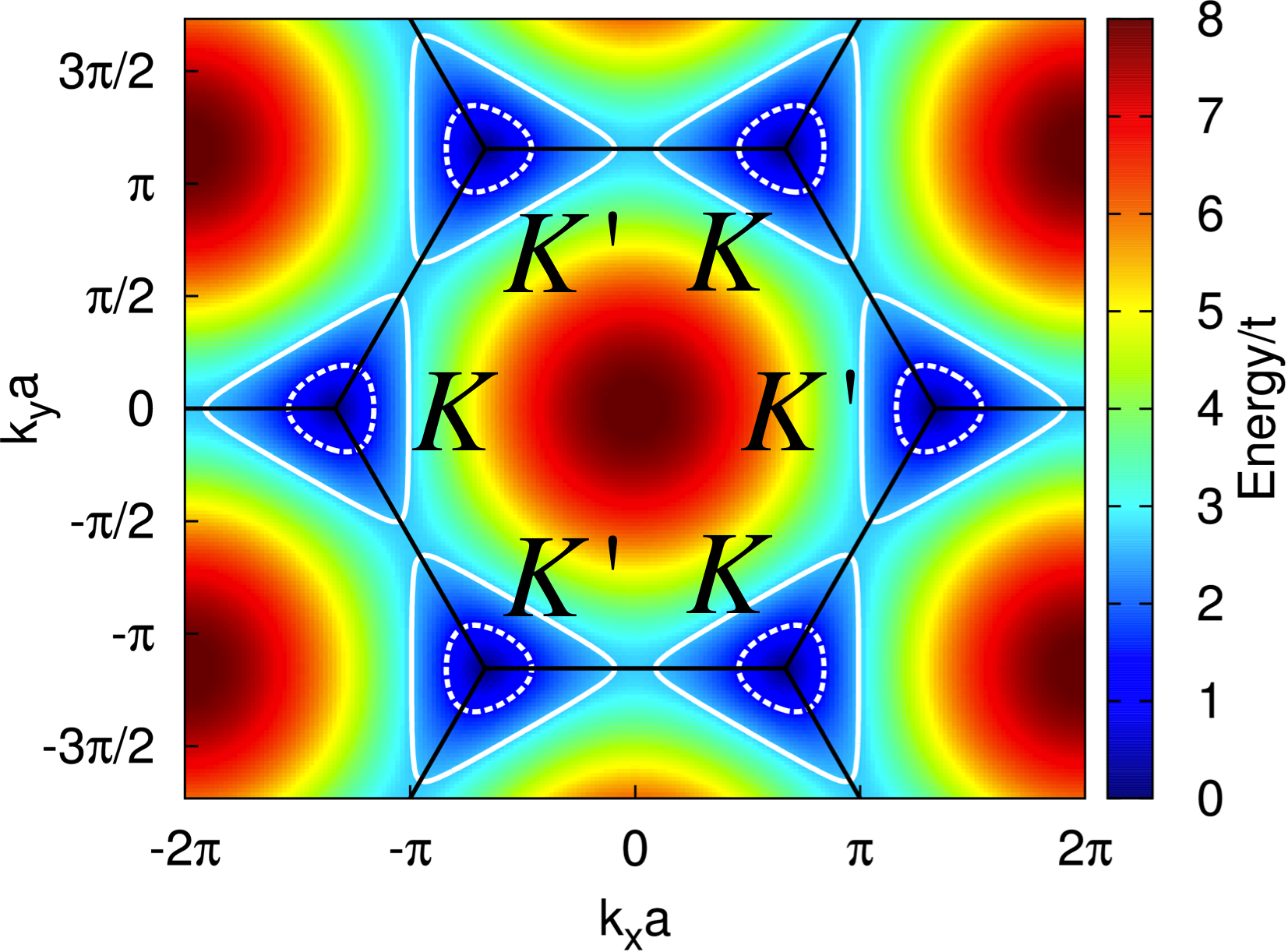}
\par\end{centering}

\caption{\label{fig:Dispersion Relation}The two-dimensional dispersion relation
for graphene demonstrates the two inequivalent valleys as cones where
the edges of the Brillouin Zones (black lines) meet. Dashed white
lines indicate the one-dimensional dispersion surface at $E=0.5t$,
while solid white lines indicate $E=0.98t$, demonstrating extreme
triagonal warping. }
\end{figure}
The band structure for graphene prominently features the two inequivalent
$\valley$ and $\othervalley$ valleys in the energy range of $-t\leq E\leq t$\citep{castro-neto},
as can be seen in Fig.~\ref{fig:Dispersion Relation}. At energies
close to the Dirac point $E=0$, these valleys exhibit a linear dispersion
relation and the electron behaves as a four-component spinor Dirac
particle (two pseudo-spins, and two traditional spins). Using the
creation operators $\mathbf{a}^{\dagger}$ and $\mathbf{b}^{\dagger}$
on the A- and B-sublattices respectively (see Fig.~\ref{fig:Edge-Close-Up}),
the two pseudospinors can be written as
\begin{eqnarray}
\psi_{\pm,\vec K}(\vec k) & = & \frac{1}{\sqrt{2}}\left(e^{-i\theta_{\vec k}/2}\mathbf{a}^{\dagger}\pm e^{i\theta_{\vec k}/2}\mathbf{b}^{\dagger}\right)\\
\psi_{\pm,\vec K'}(\vec k) & = & \frac{1}{\sqrt{2}}\left(e^{i\theta_{\vec k}/2}\mathbf{a}^{\dagger}\pm e^{-i\theta_{\vec k}/2}\mathbf{b}^{\dagger}\right),
\end{eqnarray}
where $\theta_{\vec k}=\arctan\left(\frac{q_{x}}{q_{y}}\right)$,
$\vec q=\vec k-\vec K^{(\prime)}$ and the $\pm$ signs indicate whether
the positive- or negative-energy solutions are being used\citep{castro-neto}.
While the linear dispersion no longer applies at energies above $\sim0.4t$,
the Dirac basis remains useful as a means of describing the classical
dynamics of graphene throughout the energy range $-t\leq E\leq t$.
States near the Dirac point and at the upper edge of this spectrum
are examined in this paper.%

It might be tempting to obtain a representation of either valley in
a graphene wavefunction by subtracting off a plane wave whose wavevector
corresponds to the origin of either $K$ or $K^{\prime}$ valley,
leaving behind the residual $\vec q=\vec k-\vec K^{(\prime)}$. However,
this approach only works when quasiparticles are present in only one
valley, an assumption that cannot be generally guaranteed.

On the other hand, since wavevectors for each valley are sufficiently
separated in $k$-space, the Husimi projection can distinguish each
valley unambiguously for most momentum uncertainties. Because the
valleys are part of the same band, a scattered quasiparticle from
one valley can emerge in the other\citep{ashcroft-and-mermin}. When
this occurs, the Husimi map shows quasiparticles in one valley funneling
into a drain, and quasiparticles in the other valley emitting from
a source at the same point, leaving behind a signature for inter-valley
scattering. 

Between $-t<E<t$, the Fermi energy contours warp from a circular
shape near the Dirac point to trigonal contours, which emphasize three
directions for each valley in the distribution of group velocities
$v_{g}=\Delta_{\vec k}E\left(\vec k\right)$. As a result, the magnitude
of the wavevector $\vec q=\vec k-\vec K^{(\prime)}$ depends on its
orientation: It is bounded above by 
\begin{equation}
q_{\text{up}}=\frac{2}{a}\cos^{-1}\left[\frac{1}{4t}\left(E+t+\sqrt{-3E^{2}-6Et+9t^{2}}\right)\right],
\end{equation}
and from below by 
\begin{equation}
q_{\text{low}}=\frac{2}{a}\cos^{-1}\left[\frac{1}{4t}\left(-E+t+\sqrt{-3E^{2}+6Et+9t^{2}}\right)\right].
\end{equation}
When characterizing the momentum uncertainty, we use the average of
these two quantities.

\section{Results}

\subsection{States Away from the Dirac Point\label{sub:Realtivistic-Scars}}

\begin{figure}
\begin{centering}

\par\end{centering}

\begin{centering}
\begin{overpic}[width=0.85\columnwidth]{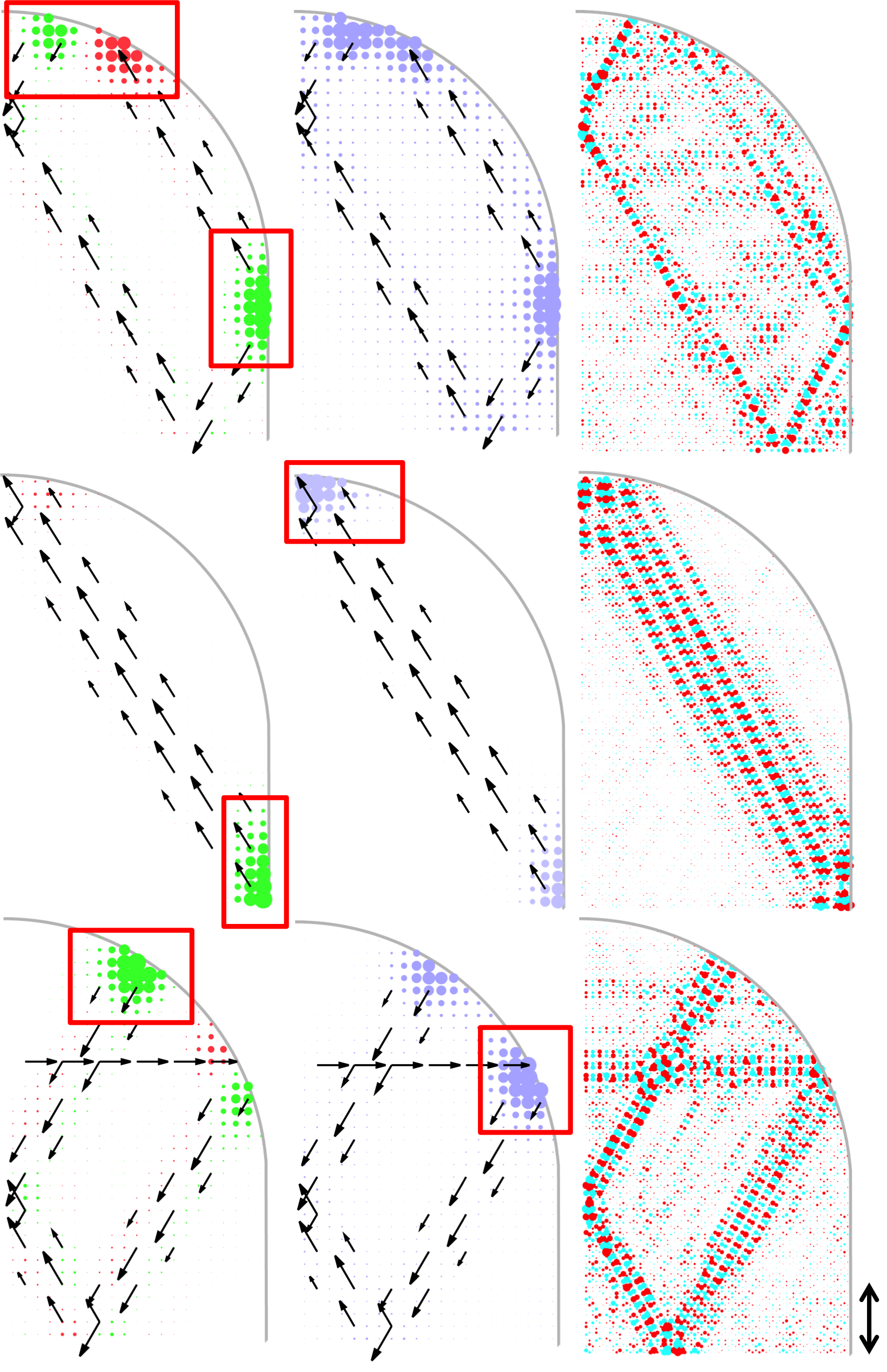}\put(-4,97){\figlab{a}}\put(-4,62){\figlab{b}}\put(-4,30){\figlab{c}}\end{overpic}
\par\end{centering}

\caption{\label{fig:Stadia}The Husimi map for three eigenstates of the closed
graphene stadium billiard with 20270 orbital sites at energy $E=0.974t\text{(a)},0.964t\text{(b)},\text{ and }0.951t\text{(c)}$.
All three calculations use coherent states with relative uncertainty
$\Delta k/k=30\%$, whose breadth is indicated by the double arrows
on the right. Only the upper-right quarter of each stadia is shown.
At left, the multi-modal analysis for the $\valley$ valley, and at
right the wavefunction representation. The divergence of the Husimi
map is indicated in green (red) to for positive (negative) values.
Angular deflection is indicated in blue (Eq.~\ref{eq:Summed-Divergence}).
Red boxes indicate the magnified views in Fig.~\ref{fig:Stadia-Zoom-In}.}
\end{figure}

\begin{figure}
\begin{centering}
\begin{overpic}[width=0.85\columnwidth]{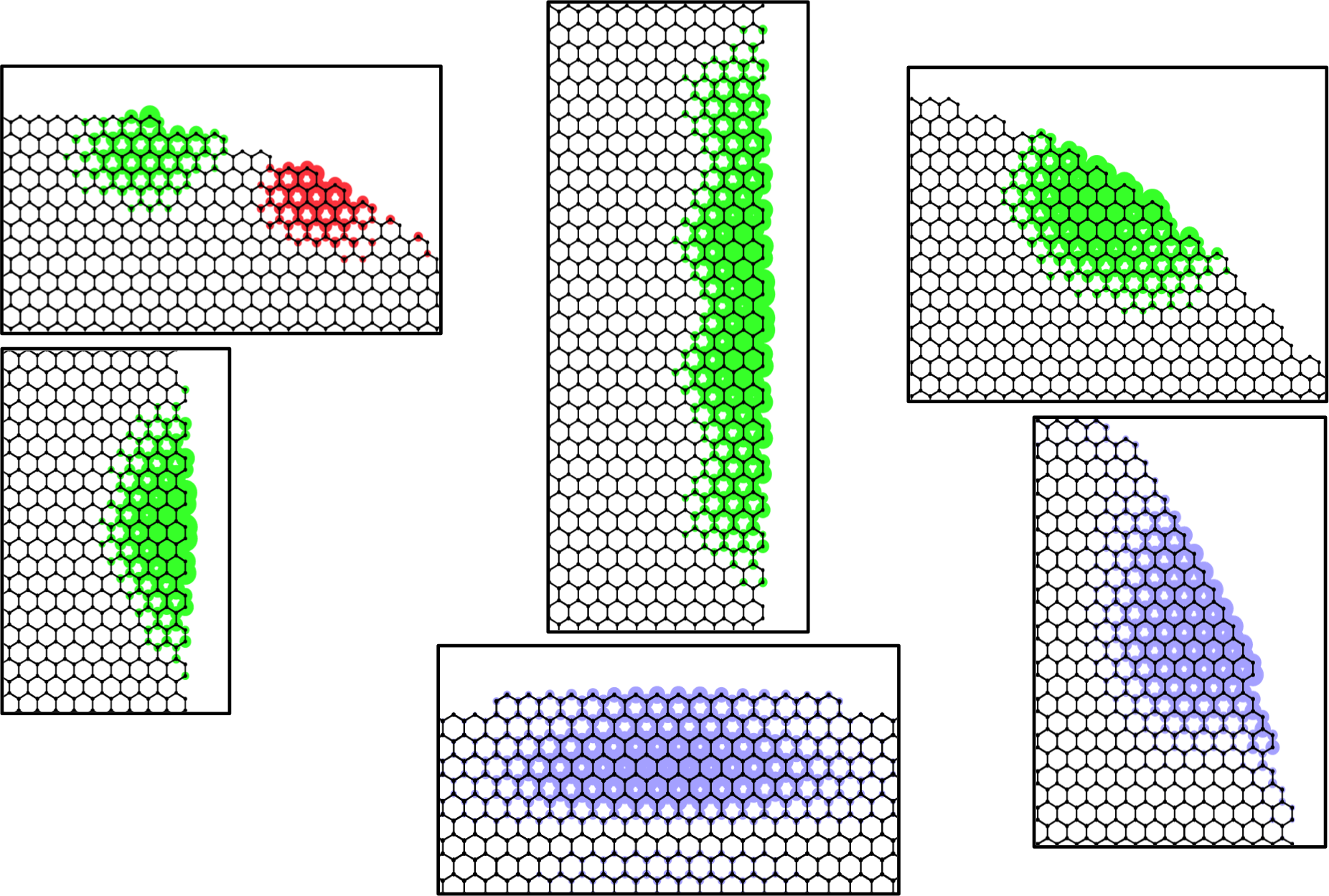}\put(0,64.5){\figlab{a}}\put(34,64.5){\figlab{b}}\put(68,64.5){\figlab{c}}\end{overpic}
\par\end{centering}

\caption{\label{fig:Stadia-Zoom-In}Magnified views of the divergence and angular
deflection in Fig.~\ref{fig:Stadia} (red boxes). The sources and
drains in the $\valley$-valley Husimi map are actually inter-valley
scattering points, which occur along non zig-zag boundaries. In contrast,
points of angular deflection that are \emph{not} sources or drains
correspond to \emph{intra-}valley scatterers and occur along pure
or nearly-pure zig-zag boundaries.}
\end{figure}

Fig.~\ref{fig:Stadia} shows Husimi maps for three eigenstates of
a large closed-system stadium billiard with 20270 orbital sites at
energies of $E=0.974t\text{(a)},0.964t\text{(b)},\text{ and }0.951t\text{(c)}$.
We have chosen these states because they exhibit very clear linear
trajectories. At energies close to $E=t$, trajectories exhibit pronounced
trigonal warping, as seen by the three preferred directions. While
the classical trajectories are obvious in the wavefunction itself,
the Husimi map identifies the direction of each trajectory with respect
to each valley.

The presence a few dominant classical paths in each wavefunction in
Fig.~\ref{fig:Stadia} allows us to infer the relationship between
boundary types and scattering among the two Dirac valleys. When a
quasiparticle in one valley scatters into the other, it appears in
the Husimi map as drain. We can measure this by summing the divergence
for all angles in the Husimi map as 
\begin{equation}
Q_{\text{div.}}\left(\vec r;\Psi\right)=\int D(\vec r,\vec k;\Psi)\left|\boldsymbol{\nabla}_{\vec k}E\left(\vec k\right)\right|d^{d}k',
\end{equation}
where $D(\vec r,\vec k;\Psi)$ is defined as the divergence of the
Husimi map for one wavevector $\vec k$,
\begin{eqnarray}
D(\vec r,\vec k;\Psi) & = & \int\sum_{i=1}^{d}\frac{\text{Hu}\left(\vec k,\vec r';\Psi\right)-\text{Hu}\left(\vec k,\vec r;\Psi\right)}{\left(\vec r'-\vec r\right)\cdot\vec e_{i}}\nonumber \\
 &  & \times\exp\left[\frac{\left(\vec r'-\vec r\right)^{2}}{2\sigma^{2}}\right]d^{d}r',\label{eq:Summed-Divergence}
\end{eqnarray}
where we sum over the $d$ orthogonal dimensions each associated with
unit vector $\vec e_{i}$. The divergence in the $\valley$ valley,
seen in green and red (for positive and negative values, respectively)
in Figs.~\ref{fig:Stadia} and \ref{fig:Stadia-Zoom-In}, shows that
the scattering points all lie along non-zig-zag boundaries. Plots
for the $\othervalley$ valley (not shown) are inverted, corroborating
the time-reversal symmetry relationship between the two valleys.

On the other hand, when a quasiparticle in one valley reflects off
a boundary but does \emph{not }scatter into the other valley, the
divergence is zero, but the reflection can still be measured in the
angular deflection of the Husimi map,
\begin{equation}
Q_{\text{ang.}}\left(\vec r;\Psi\right)=\int\left|D_{\text{abs.}}(\vec r,\vec k;\Psi)\boldsymbol{\nabla}_{\vec k}E\left(\vec k\right)\right|d^{d}k.
\end{equation}
 $D_{\text{abs.}}(\vec r,\vec k;\Psi)$ is defined as the \emph{absolute}
divergence of the Husimi function for one particular trajectory angle
with a wavevector $\vec k$,
\begin{eqnarray}
D_{\text{abs.}}(\vec r,\vec k;\Psi) & = & \int\sum_{i=1}^{d}\left|\frac{\text{Hu}\left(\vec k,\vec r';\Psi\right)-\text{Hu}\left(\vec k,\vec r;\Psi\right)}{\left(\vec r'-\vec r\right)\cdot\vec e_{i}}\right|\nonumber \\
 &  & \times\exp\left[\frac{\left(\vec r'-\vec r\right)^{2}}{2\sigma^{2}}\right]d^{d}r'.\label{eq:Summed-Divergence-1}
\end{eqnarray}
 As a result, boundary points with large angular deflection are either
inter-valley or intra-valley scatterers depending on the magnitude
of divergence at each point.

In Figs.~\ref{fig:Stadia} and \ref{fig:Stadia-Zoom-In}, we plot
the angular deflection in blue to compare to the divergence in green
and red. Using this information, we can determine that for the wavefunction
in Fig.~\ref{fig:Stadia}a, all boundary scattering points are inter-valley
scatterers, since all points of angular deflection exhibit divergence.
The wavefunction in Fig.~\ref{fig:Stadia}b, on the other hand, only
exhibits divergence along the vertical sides of the stadium billiard:
the horizontal top edge exhibits strong angular deflection but \emph{no
}divergence, and constitutes an \emph{intra}-valley scatterer. Examining
the magnified views in Figs.~\ref{fig:Stadia-Zoom-In}a and \ref{fig:Stadia-Zoom-In}b,
we see that inter-valley scatterers correspond to armchair edges,
and the intra-valley scatterers belong to zig-zag edges, corroborating
the findings at the Dirac point by Akhmerov and Beenakker\citep{BeenakkerBC}.
Similar points of scattering can also be found in Figs.~\ref{fig:Stadia}c
and \ref{fig:Stadia-Zoom-In}c.

Because of the time-reversal relationship between the two valleys,
the severe restriction on group velocities, and the placement of zig-zag
and armchair boundaries, no path at these energies exists \emph{without}
interacting with an inter-valley scatterer (data not shown). By comparison,
it is not only possible but common to find states near the Dirac point
that exhibit the opposite: all boundary conditions which are expressed
belong to only \emph{intra}-valley scatterers (See Sec.~\ref{sub:Low-Energy-States}). 

\begin{figure}
\begin{centering}
\includegraphics[width=0.95\columnwidth]{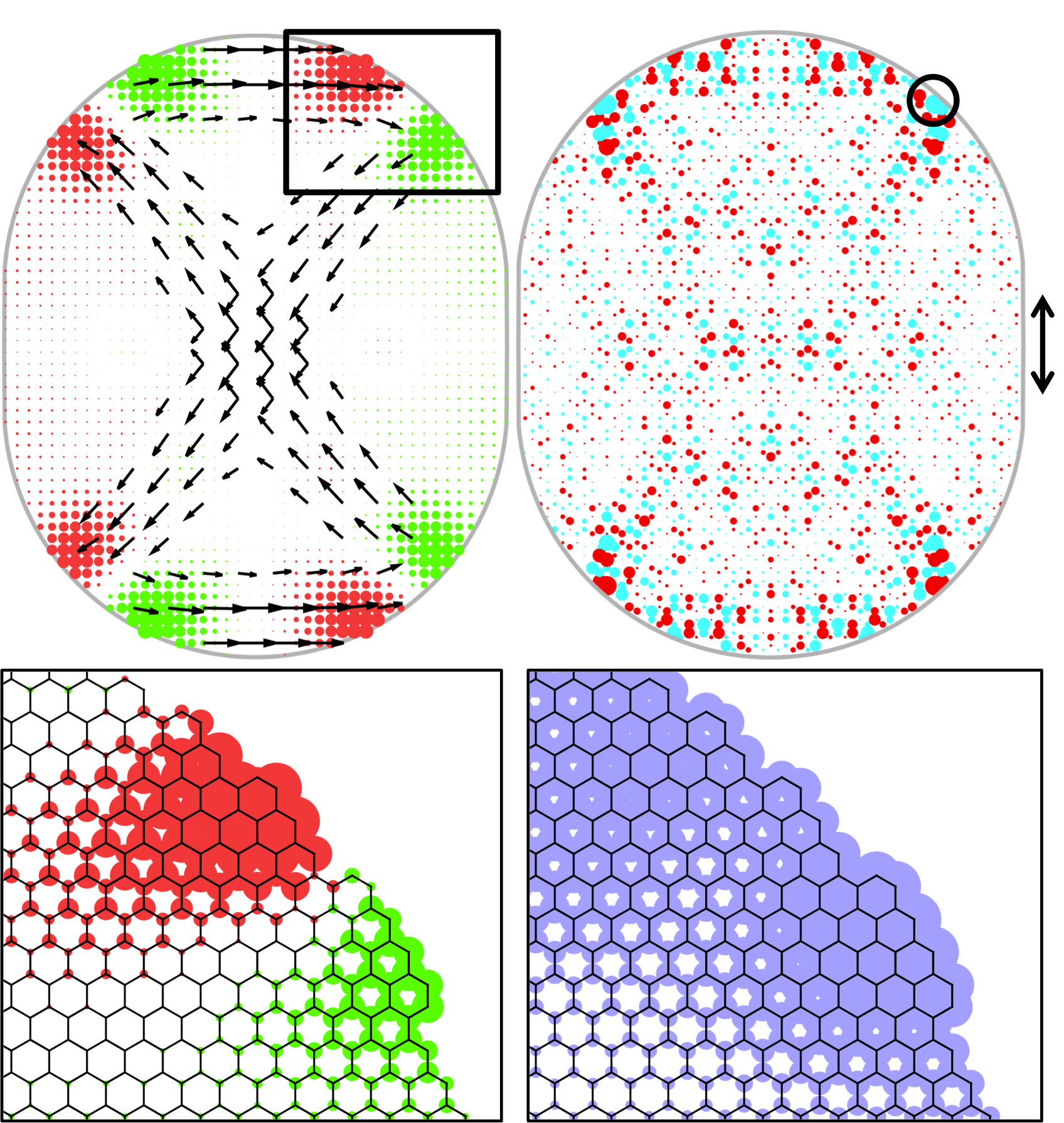}
\par\end{centering}

\caption{\label{fig:Resonant_Eigenstate}A closed-system eigenstate at $E=0.72t$
for the smaller graphene stadium. At top, the filtered Multi-Modal
analysis is with relative momentum uncertainty $\Delta k/k=30\%$
along with the wavefunction (right). The spread of the coherent state
is indicated by the double arrows. At bottom, higher-resolution calculations
of the divergence (green for positive, red for negative) and the angular
deflection (blue) are shown against the graphene structure. The black
circle indicates where the system boundary is perturbed in the original
paper\citep{Huang-PRL-Scars} as discussed in Section \ref{sub:Fano-Resonance}.}
\end{figure}

In comparison to Fig.~\ref{fig:Stadia}, the eigenstate of the much
smaller graphene stadium system in Fig.~\ref{fig:Resonant_Eigenstate}
does not appear to show isolated trajectories in its wavefunction
representation. This is not surprising since this system can only
accommodate five deBroglie wavelengths vertically, and three horizontally,
severely restricting its ability to resolve such trajectories. However,
clear self-retracing trajectories are quite visible in the Husimi
map in Figs.~\ref{fig:Resonant_Eigenstate}, with evident sources
and drains inhabiting the boundary, showing that the Husimi map can
yield a semiclassical interpretation of the dynamics of the states
not possible from just the wavefunction of the system. Moreover, because
the paths indicated by the Husimi map marshal the electron away from
lateral boundaries, where leads connect to produce the open system
in Sec.~\ref{sub:Fano-Resonance}, the Husimi map helps us understand
the role this state plays in forming a long-lived resonance in the
open system.

In both Figs.~\ref{fig:Stadia} and \ref{fig:Resonant_Eigenstate},
wavefunctions in graphene away from the Dirac point are linked to
valley switching classical ray paths which bounce back and forth along
straight lines. These wavefunction enhancements are not strictly scars\citep{PhysRevLett.53.1515},
as first suggested by Huang \emph{et al.}\citep{Huang-PRL-Scars},
since scars are generated by unstable classical periodic orbits in
the analogous classical limit (group velocity) system. Instead, the
wavefunction structures are more likely normal quantum confinement
to stable zones in classical phase space constrained by group-velocity
warping at these energies.

\subsection{States Near the Dirac Point\label{sub:Low-Energy-States}}

\begin{figure}
\begin{centering}
\includegraphics[width=0.85\columnwidth]{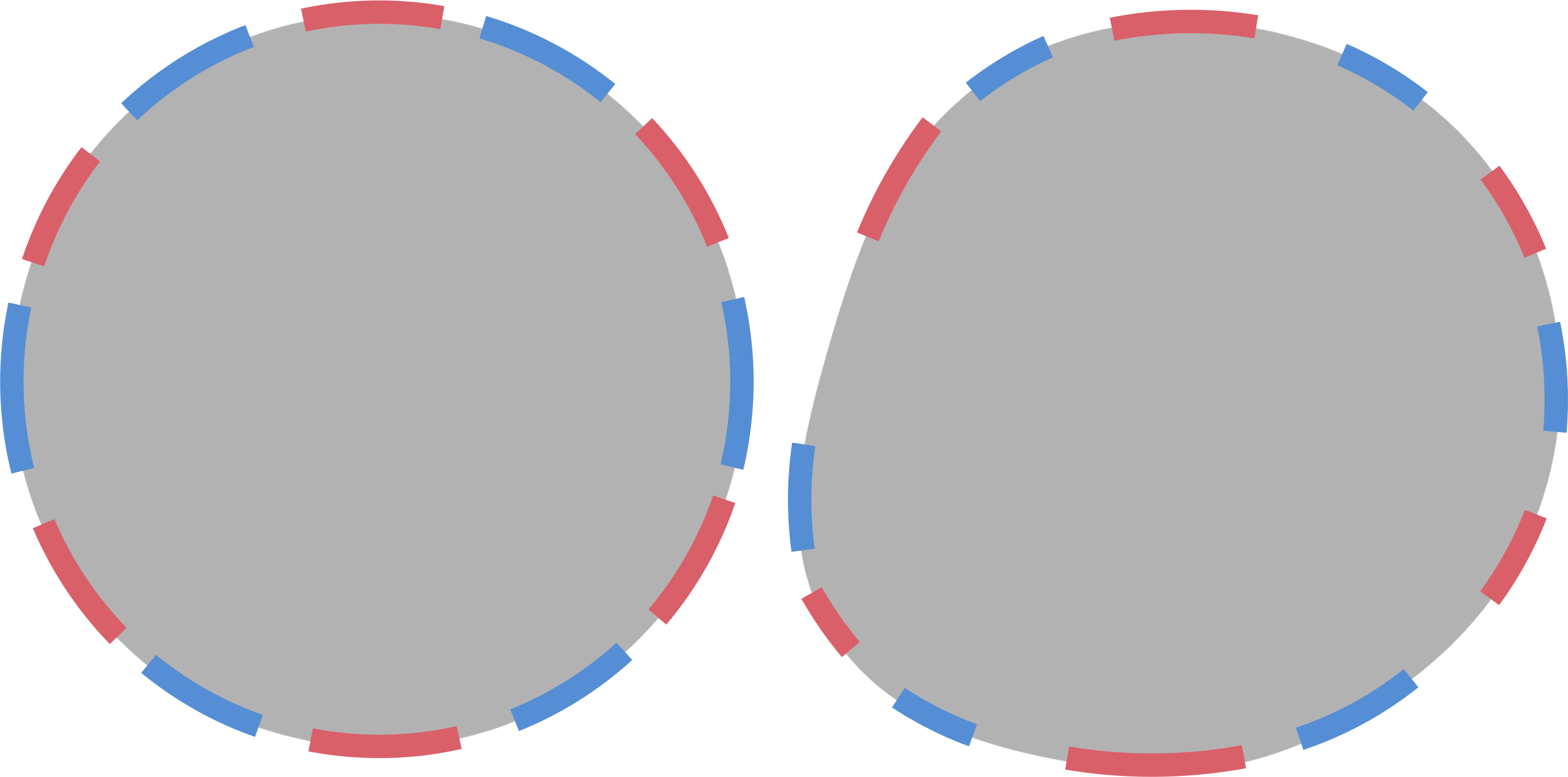}
\par\end{centering}

\caption{\label{fig:Circle Schematic}Schematic indicating the locations of
armchair (blue) and zig-zag (red) edges in the circular system (left)
and the Wimmer system (right).}
\end{figure}

\begin{figure}
\begin{centering}
\begin{overpic}[width=1.0\columnwidth]{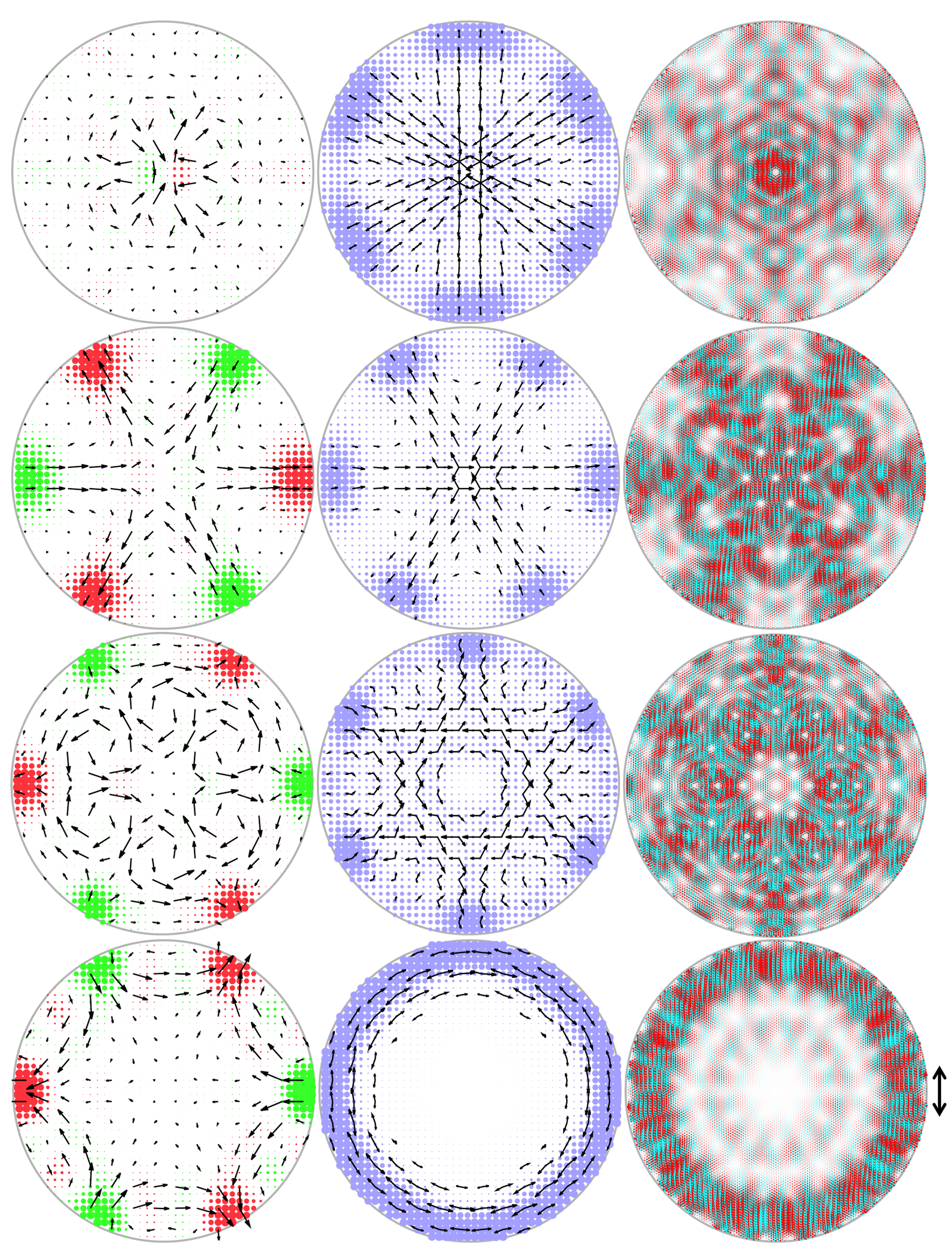}\put(1,96.5){\figlab{a}}\put(1,73){\figlab{b}}\put(1,48){\figlab{c}}\put(1,23.5){\figlab{d}}\end{overpic}
\par\end{centering}

\caption{\label{fig:Four-Circles}Low-energy graphene states require additional
tools to fully grasp the classical dynamics. The Husimi map for the
$\valley$-valley is plotted for four eigenstates of a closed circular
system with 71934 orbital sites at energies around $E=0.2t$. All
three calculations use coherent states with relative uncertainty $\Delta k/k=20\%$,
with breadth indicated by the double arrows on the right. From left-to-right:
the Husimi flux, multi-modal analysis, and the wavefunction. The divergence
of the Husimi flux is indicated in green (red) for positive (negative)
values. In blue, the angular deflection. }
\end{figure}
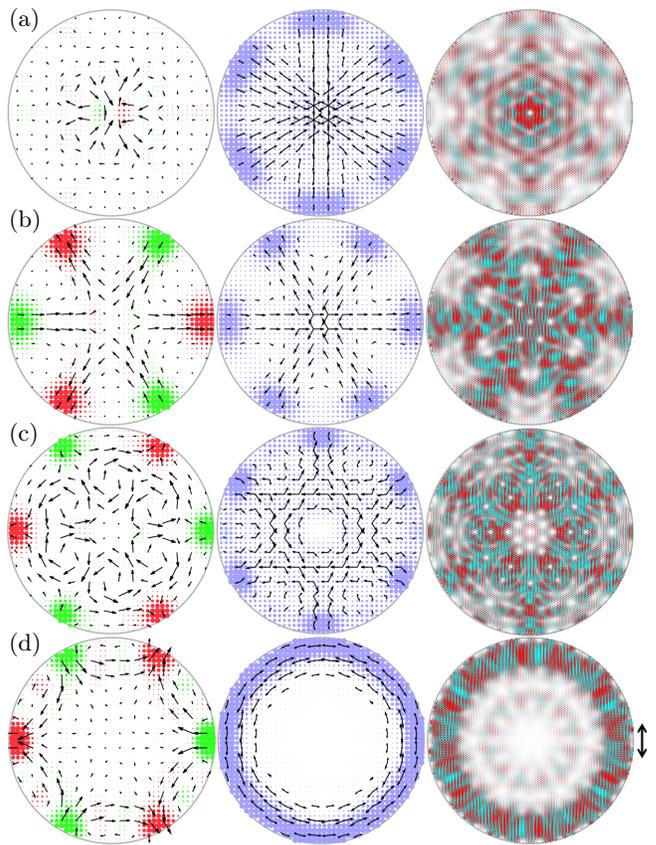

We now explore the properties of low-energy closed-system states in
graphene, using the circular graphene flake and the distorted circular
flake introduced by Wimmer \emph{et al.}\citep{Wimmer-Robusteness}.
The latter geometry was chosen because its dynamics are chaotic and
sensitive to the placement of armchair and zig-zag boundaries, which
shift as a result of the distortion. We indicate the two boundary
types for both geometries in Fig.~\ref{fig:Circle Schematic}..

In the continuous system, the Fermi wavevector grows with the square-root
of the energy, but in graphene, the effective wavevector $\vec q=\vec k-\vec K^{(\prime)}$
grows linearly. As a result, the deBroglie wavelength is much larger
for the graphene system than for the continuous system at similar
energy scales, making it difficult to conduct calculations with sufficient
structure in the wavefunction. Consequently, we examine states at
energies away from the Dirac point to bring calculations within a
reasonable scope. (For instance, we have selected a system size under
100,000 orbital sites to facilitate replication of our results). Since
trigonal warping becomes significant above $E=0.4t$, we have selected
the energy of $0.2t$ for all states in our analysis to maximize the
number of wavelengths within a small graphene system while maintaining
the same physics from energies closer to the Dirac point.

Fig.~\ref{fig:Four-Circles} shows four eigenstates of the circular
graphene flake. Like the free-particle circular well, eigenstates
of the graphene circular flake resemble eigenstates of the angular
momentum operator (see Mason \emph{et al.}\citep{Mason-Husimi-Continuous}
for direct comparisons and Husimi maps). For instance, the wavefunctions
in Figs.~\ref{fig:Four-Circles}a-b are radial-dominant, while the
wavefunction in Fig.~\ref{fig:Four-Circles}d is angular-dominant.
These observations carry over to the dynamics of the wavefunctions
revealed by the multi-modal analysis for the $\valley$ valley, which
shows radially-oriented paths in Figs.~\ref{fig:Four-Circles}a-b
and circular paths skimming the boundary in Fig.~\ref{fig:Four-Circles}d.
Fig.~\ref{fig:Four-Circles}c shows a state with a mixture of radial
and angular components; in the multi-modal analysis, this appears
as straight paths between boundary points highlighted by the angular
deflection.

Unlike free-particle circular wells, however, the lattice sampling
on the honeycomb lattice breaks circular symmetry and replaces it
with six-fold symmetry. Because eigenstates of the system emphasize
certain boundary conditions, the manner in which each state establishes
itself strongly varies. For instance, the two radial-dominant states
in Figs.~\ref{fig:Four-Circles}a-b exhibit intravalley (a) or intervalley
(b) scattering. Accordingly, the locations where the rays terminate
on the boundary correlate with zig-zag and armchair boundaries respectively.
The wider spread in angular deflection in Fig.~\ref{fig:Four-Circles}a
corroborates Akhmerov and Beenakker\citep{BeenakkerBC}, showing that
that intravalley scattering occurs over a larger set of boundaries
than intervalley scattering.

Because each valley reflects back to itself in Fig.~\ref{fig:Four-Circles}a,
there is no net flow of either valley in the bulk of the system. As
a result, the multi-modal analysis shows counter-propagating flows,
and the Husimi flux (Eq.~\ref{eq:Husimi-Vector}) is zero except
at the center, where slight offsets in trajectories form characteristic
vortices. In Fig.~\ref{fig:Four-Circles}b, on the other hand, each
ray in the wavefunction is associated with a distinct source and drain,
which is evident in both the multi-modal analysis and the Husimi flux.

In Figs.~\ref{fig:Four-Circles}c-d, the locations of sources and
drains for the $\valley$ valley are reversed from Fig.~\ref{fig:Four-Circles}b.
However, the roles that inter-valley scattering play in these states
is less clear; rather, inter- \emph{and }intra-valley scattering dominate
these wavefunctions. In Fig.~\ref{fig:Four-Circles}c, this can been
by the emphasis of angular deflection along the zig-zag boundaries,
which do not show any divergence. In Fig.~\ref{fig:Four-Circles}d,
even though the wavefunction and the multi-modal analysis clearly
emphasize a classical path that skims the boundary, the path actually
flips between each valley each time it encounters an inter-valley
scatterer. For both states, the various trajectories merge to form
vortices in the Husimi flux, with sources and drains at armchair edges.

\begin{figure}
\begin{centering}
\begin{overpic}[width=1.0\columnwidth]{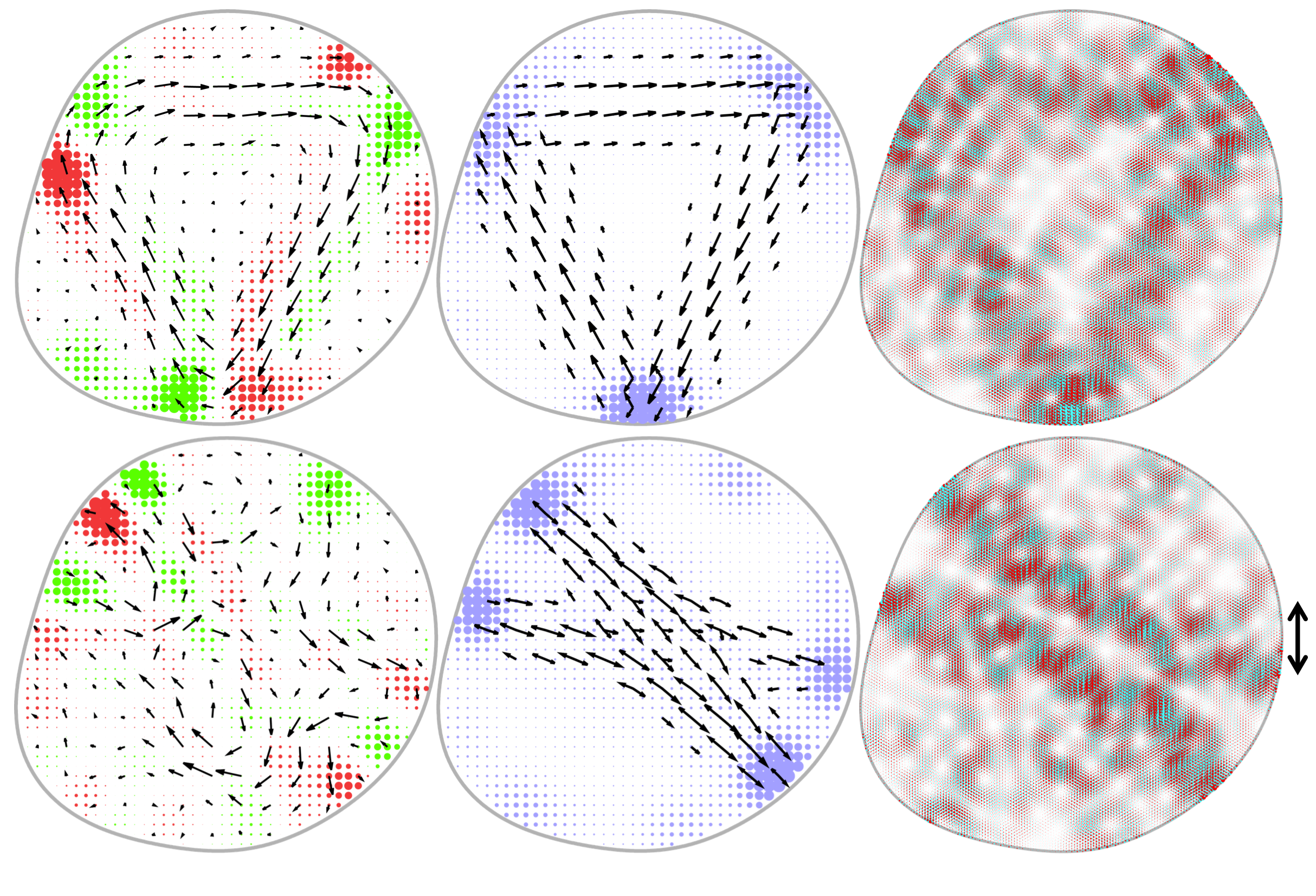}\put(1,66){\figlab{a}}\put(1,33){\figlab{b}}\end{overpic}
\par\end{centering}

\caption{\label{fig:Wimmer}In parts (a) and (b), the same information is plotted
as in Fig.~\ref{fig:Four-Circles}, but for the Wimmer system (see
Fig.~\ref{fig:Circle Schematic}), with 96425 orbital sites. These
states also have energies near $E=0.2t$ and are represented by coherent
states of uncertainty $\Delta k/k=20\%$ with breadth indicated by
the double arrows.}
\end{figure}
When the circular flake is distorted, as in the Wimmer system (Figs.~\ref{fig:Circle Schematic}
and \ref{fig:Wimmer}), inter- and intra-valley scatterers are re-arranged
and re-sized as a function of the local radius of curvature of the
boundary.

Figs.~\ref{fig:Wimmer}a-b show two eigenstates of the Wimmer system.
The boundary conditions for these states most closely resemble Fig.~\ref{fig:Resonant_Eigenstate},
since sources and drains appear next to each other. This is a signature
of mixed scattering -- both inter- and intra-valley scattering occur
in various proportions at these points. For example, the multi-modal
analysis in Fig.~\ref{fig:Wimmer}a shows a triangular path, but
not all legs of the triangle are equally strong, corresponding to
various degrees of absorption and reflection at each scattering point
which can be seen in the divergence.

\begin{figure}
\begin{centering}
\includegraphics[width=0.85\columnwidth]{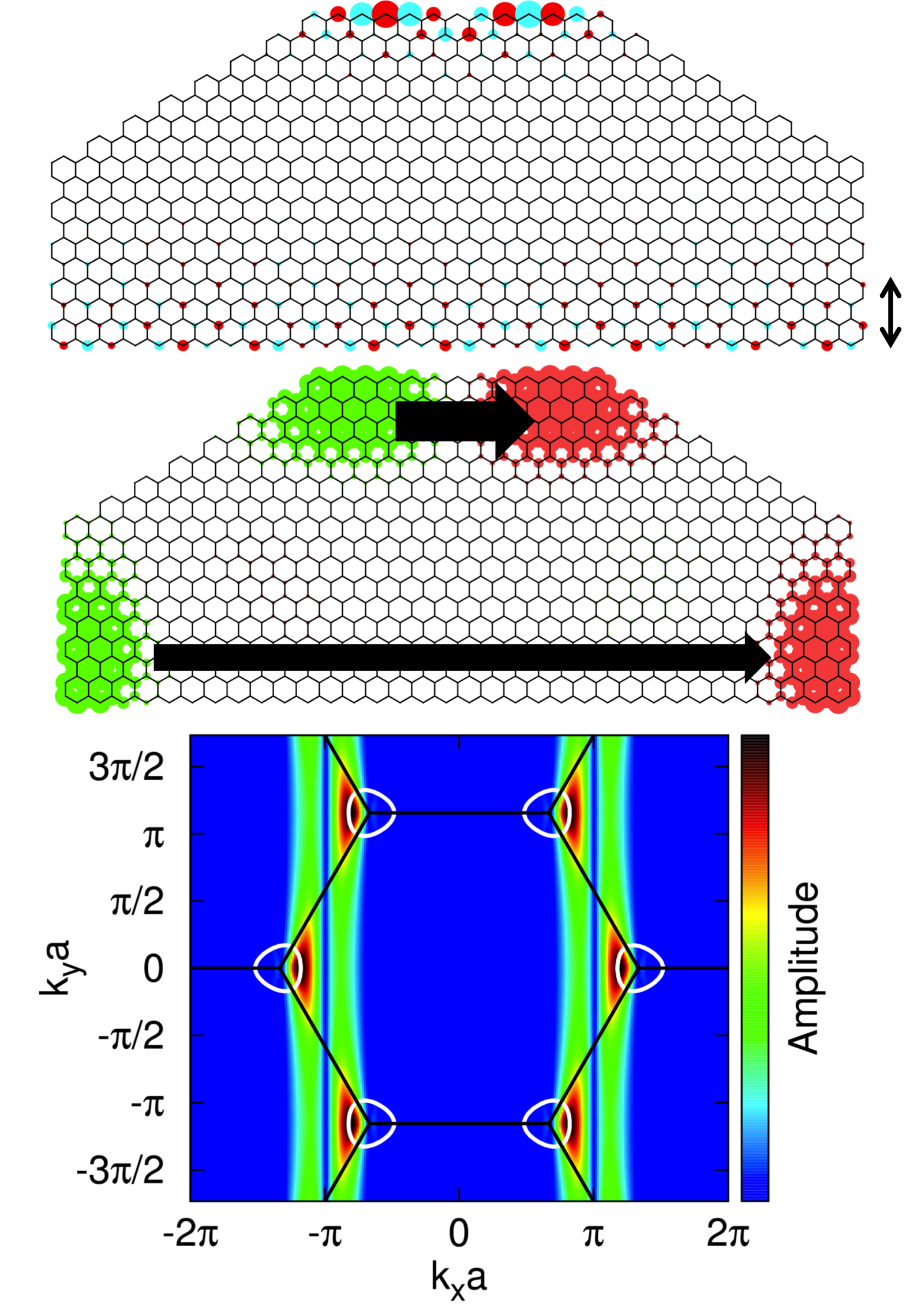}
\par\end{centering}

\caption{\label{fig:Edge State}An extremely small ``rooftop'' graphene flake
at energy $E=0.0015735t$ showing two edge states at the top and bottom
boundaries which tunnel into each other. At top, the full wavefunction,
at middle, divergence is indicated in green and red, and a schematic
of the Husimi flux for the $\valley$ valley is shown. At bottom,
the Fourier transform of the state is shown, with the contour line
used to generate the Husimi map in white, set at an arbitrary energy
in order to maximize the intersection of the contour with the Fourier-transform
amplitude. The spread of the wavepacket used to generate this map
is indicated by the double arrows.}
\end{figure}
Edge-states are a set of zero-energy surface states that are strongly
localized to zig-zag boundaries and potentially long-lived\citep{castro-neto},
and since they can be used as modes of transport\citep{Wimmer-Spin-Currents-Rough-Nanoribbons,Wimmer-Spin-Transport}
and be strongly spin-polarized\citep{KaxirasNanoFlakes,WeiLi}, they
have been proposed a candidate for spin-tronics devices\citep{castro-neto,KaxirasNanoFlakes,Wimmer-Spin-Currents-Rough-Nanoribbons,Wimmer-Spin-Transport,WeiLi}.
But because edge states exhibit a different dispersion relation than
the two valleys in the bulk, they cannot be ``sensed'' by the $\valley$
or $\othervalley$ valley Husimi projections. Instead, the Husimi
map can be generated using wavevectors appropriate to the edge states,
which shows them as standing waves on the surface (see Fig.~\ref{fig:Edge State}).
As noted in Wimmer \emph{et al.}\citep{Wimmer-Robusteness}, it is
possible for edge states to tunnel into each other using bulk states
as a medium, but we have found that $\valley$ or $\othervalley$
valley Husimi maps of bulk states which hybridize with them are indistinguishable
from their non-hybridized counterparts

\subsection{Fano Resonance\label{sub:Fano-Resonance}}

\begin{figure}
\begin{centering}
\includegraphics[width=0.95\columnwidth]{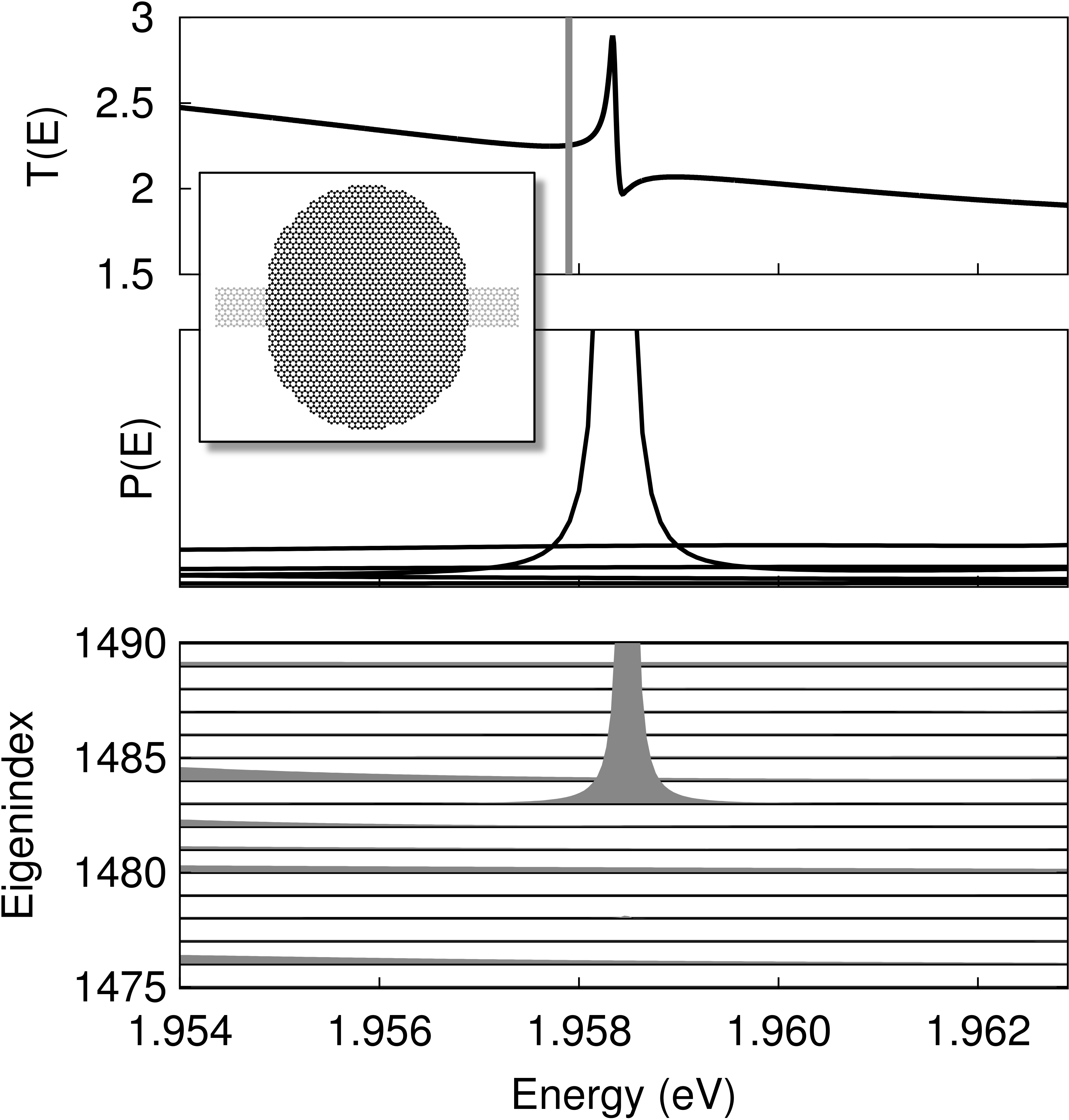}
\par\end{centering}

\caption{\label{fig:Fano}System properties of the scattering density matrix
$\rho$ around the Fano resonance centered at $E=1.9582\text{eV}$
for the open system in the inset. Top: The transmission profile across
the two leads, with the closed-system eigenstate energy at $E=1.9579\text{eV}$,
corresponding to the eigenstate at index 1483 (below), indicated by
the vertical grey line. Middle: Diagonalizing the density matrix produces
a handful of non-trivial scattering wavefunctions in its eigenvectors.
The eigenvalues of these vectors, which correspond to their measurement
probability, are graphed. The wavefunction associated with the closed-system
eigenstate hybridizing with the direct channel peaks strongly around
the Fano resonance. Bottom: The density matrix is projected onto the
closed-system eigenstates, showing that eigenstate 1483 strongly peaks
at the Fano resonance. }
\end{figure}

This section addresses Fano resonance\citep{fano-original} in graphene
systems, a conductance phenomenon that occurs as a result of interference
between a direct state (conductance channel) and a quasi-bound indirect
state similar to the eigenstates this paper has examined. Fano resonances
are an ideal case study for the Husimi map, not only because they
are ubiquitous in theory\citep{ferry-pointer-states,Munoz-Rojas2006}
and experiments\citep{Fano-Experiment2,Fano-Experiment,Fano-Experiment3},
but also because their behavior is well-understood\citep{datta,mesoscopicbook,ferrybook,transport-book2,dots-transport-book,datta2,fano-review}%
. However, Fano resonances in \emph{graphene} quantum dots are less
well characterized\citep{Wimmer-Energy-Levels,Huang-Trans-Scar,Huang-Open-Dots,Huang-Cond-Fluc}
and lack a comprehensive theory relating boundary conditions to bulk
state behavior in graphene.

To study Fano resonance, we first compute a scattering wavefunction
using the recursive numerical Green's function method described in
Mason \emph{et al.}\citep{mason}. This method produces a scattering
density matrix $\rho$, which is diagonalized. Each eigenvector corresponds
to a scattering wavefunction, which has an associated eigenvalue indicating
its measurement probability (Fig.~\ref{fig:Fano}, middle). We focus
on the resonance study in Huang\emph{ et al.}\citep{Huang-PRL-Scars}.

The resonance in Fig.~\ref{fig:Fano} is associated with the eigenstate
from Fig.~\ref{fig:Resonant_Eigenstate} of the closed billiard system.
This eigenstate couples only weakly to leads which are attached at
its sides (shown in the inset of Fig.~\ref{fig:Fano}). This makes
it possible for a scattering electron to enter the system through
a direct channel but then become trapped in a quasi-bound state related
to the eigenstate, causing the density of states projected onto the
eigenstate to strongly peak near its eigenenergy (Fig.~\ref{fig:Fano},
bottom). As the system energy sweeps across the eigenenergy, the phase
of the eigenstate component shifts through $\pi$, causing it to interfere
negatively and then positively with the direct channel, giving rise
to the distinctive Fano curve (Fig.~\ref{fig:Fano}, top). As a result,
the scattering wavefunction with the largest measurement probability
is in fact a hybridized state between the closed-system eigenstate
and the direct channel, and its probability peaks around an energy
near, but not exactly the same as, the eigenstate energy (Fig.~\ref{fig:Fano},
middle). The shift in energy arises as a perturbation from the leads.

For closed graphene systems, the two valleys satisfy time-reversal
symmetry as an analytical consequence of lattice sampling on the honeycomb
lattice. As a result, trajectories in one valley are exactly reversed
from the other valley, in analogy with free-particle systems where
opposing trajectories cancel each other produce zero flux. This observation
allows us to remove the time-reversal symmetry of a scattering wavefunction
by summing the projections for both valleys, revealing the time-reversal
\emph{asymmetric }part of the wavefunction.

\begin{figure}
\begin{centering}
\includegraphics[width=0.85\columnwidth]{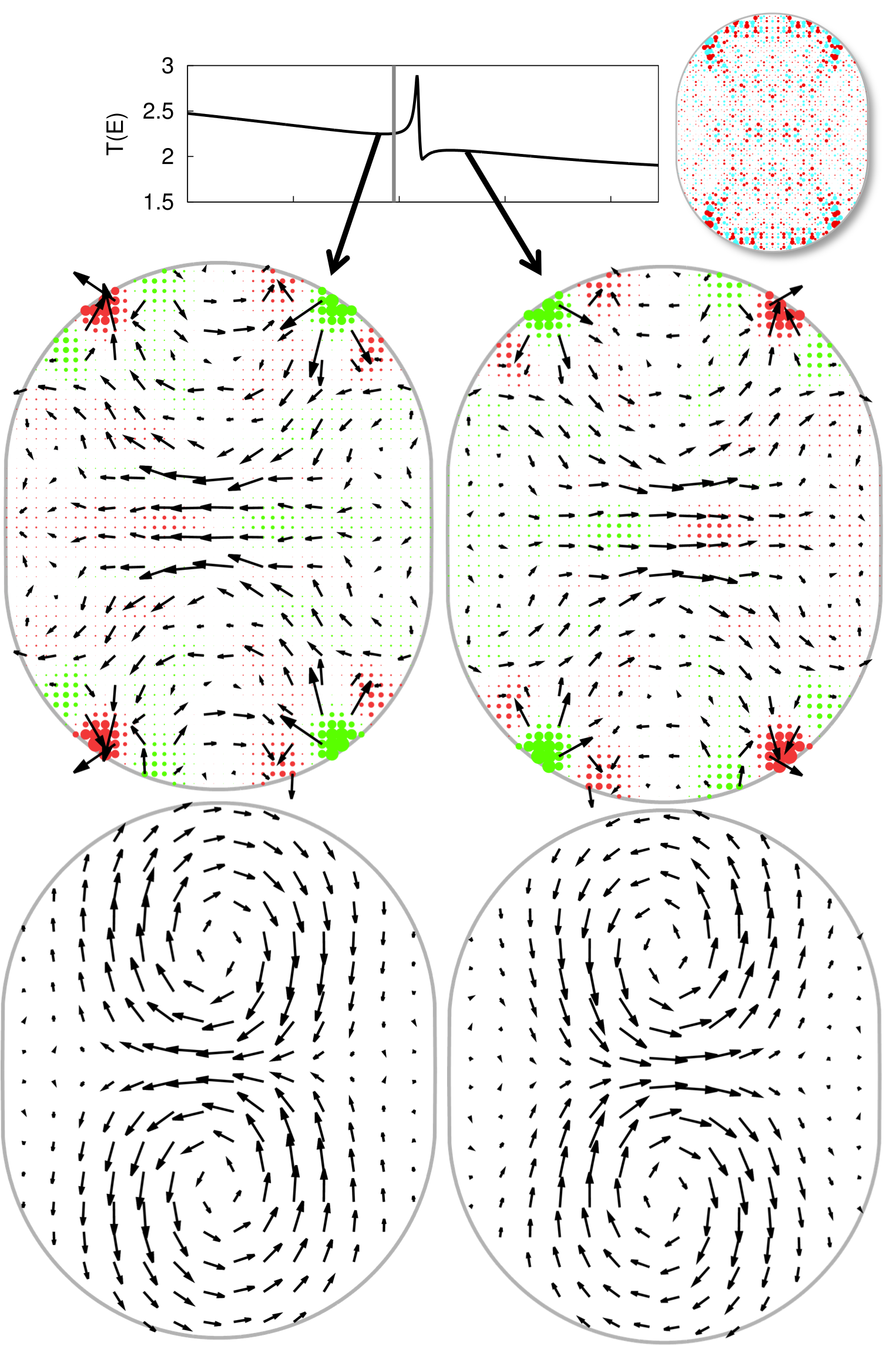}
\par\end{centering}

\caption{\label{fig:Above-at-below}Above and below the Fano resonance in Figs.~\ref{fig:Fano}
(inset), the time-reversal symmetry between the $K$ and $K'$ valleys
is lifted, making it possible to add the Husimi flux for both valleys
to measure valley-polarized current. Above, the Husimi flux maps of
both valleys are added for the scattering wavefunction at energies
$E=1.9582t$ and $1.9586t$, with $\Delta k/k=30\%$. Below, the probability
flux, convolved with a Gaussian kernel of the same size as the coherent
state. At energies this close to resonance, the wavefunction does
not visually change from the closed-system eigenstate in the inset,
but the residual current that occurs near these resonances switches
direction across resonance. }
\end{figure}

Fig.~\ref{fig:Above-at-below} shows the results of adding the Husimi
flux maps of both valleys at two energies, below and above resonance.
We find sources and drains in the summed Husimi flux map at the corners
of the system where the classical paths of the $K'$-valley Husimi
map (Fig.~\ref{fig:Resonant_Eigenstate}) reflect off the system
boundary.

To understand why, we consider that during transmission, quasiparticles
enter from the left incoming lead and exit through the right outgoing
lead. However, near resonance, the wavefunction is strongly weighted
by the closed-system eigenstate, which has \emph{no} net quasiparticle
current. Husimi maps for either valley also reflect this fact: they
are indistinguishable from the Husimi maps of the closed-system eigenstate
in Fig.~\ref{fig:Resonant_Eigenstate}, and the two valleys are inverse
images of each other.

But the Husimi maps for the two valleys don't exactly cancel each
other out. When we add them together to reveal the time-reversal asymmetric
behavior of the wavefunction, the residual shows sources and drains
of net quasiparticle flow which are strongly related to the Husimi
maps for each valley, and do \emph{not} show left-to-right transmission.
Instead, the summed Husimi flux map shows the influence of transmission
on the strongly-emphasized classical paths underlying the closed-system
eigenstate. 

To compare the summed Husimi flux map to the traditional flux, we
consider the probability flow between two adjacent carbon atom sites
called the bond current, defined as 
\begin{equation}
j_{i\rightarrow j}=\frac{4e}{h}\text{Im}\left[H_{ij}G_{ij}^{n}(E)\right],
\end{equation}
where $H_{ij}$ and $G_{ij}^{n}(E)$ are the off-diagonal components
of the Hamiltonian and the electron correlation function between orbital
sites $i$ and $j$\citep{datta,graphene-bond-current2}. The electron
correlation function is proportional to the density matrix, but in
our calculations, we examine just one scattering state, so that $G_{ij}^{n}\propto\psi_{i}\psi_{j}^{\ast}$
where $\psi_{i}$ is the scattering state probability amplitude at
orbital site $i$. We can obtain a finite-difference analog of the
continuum flux operator by defining
\begin{equation}
\vec j_{i}=\sum_{j}j_{i\rightarrow j}\frac{\vec r_{j}-\vec r_{i}}{\left|\vec r_{j}-\vec r_{i}\right|^{2}},\label{eq:finite-diff-flux}
\end{equation}
which computes the vector sum of each bond current associated with
a given orbital\citep{graphene-bond-current}. 

Convolving the flux defined in Eq.~\ref{eq:finite-diff-flux} with
a Gaussian kernel of the same spread as the coherent state used to
generate the Husimi map creates an analog to the Husimi flux, except
that the convolved flux does not distinguish among valleys. We show
the convolved flux at the bottom of Fig.~\ref{fig:Above-at-below},
and find that it forms vortices which correlates with the summed Husimi
flux maps, and also fails to show the left-to-right flow responsible
for transmission. 

This behavior is directly analogous to flux in continuum systems,
where flux vortices above and below resonance show local variations
of flow but not the left-to-right drift velocity responsible for transmission.
We can recover the left-to-right flow only by examining the system
at larger scales using larger Gaussian spreads (not shown)\citep{Mason-Husimi-Continuous}.
Because of the $\pi$ phase shift of the indirect channel across resonance,
local flows reverse direction above and below resonance, but they
do not affect the left-to-right flow at larger scales except exactly
on resonance.

The stable orbits that underly the indirect channel, shown in Fig.~\ref{fig:Resonant_Eigenstate},
can be dramatically disturbed by slight modifications of the boundary
where the classical paths reflect off the boundary. The original authors
Huang \emph{et al.}\citep{Huang-PRL-Scars} examined the relationship
between system symmetry and strength of the Fano resonances by slightly
modifying the system boundary at the black circle in Fig.~\ref{fig:Resonant_Eigenstate},
and demonstrated that some resonances were drastically reduced by
this modification. 

We have chosen the resonance in this study because the Fano resonance
profile associated with it was among the most-reduced as a result
of their system modification, and our analysis provides a clear picture
as to why: the system is perturbed precisely at the boundary where
the eigenstate in Fig.~\ref{fig:Resonant_Eigenstate} has the largest
probability amplitude. With the semiclassical picture, we are able
to add to this finding an intuitive understanding: by disturbing the
reflection angle at the exact point where the two valleys scatter,
each time the electron scatters off that point some of its probability
leaves the stable orbit. The authors effectively introduced a leak
into the orbit, reducing its lifetime and the strength of its resonance
considerably. %

\section{Conclusions}

We have examined the semiclassical behavior of graphene systems using
a generalized technique that produces a vector field from projections
onto coherent states, forming an infinitely tunable bridge between
the large-scale Dirac effective field theory and the underlying atomistic
model\citep{castro-neto}. We have used this technique, called the
Husimi map, to examine the relationship between graphene boundary
types and the classical dynamics of quasiparticles in each valley
of the honeycomb dispersion relation, looking at states with energies
both close to and far from the Dirac point. We have shown that closed-system
eigenstates are associated with valley-polarized currents with zero
net quasiparticle production. We show that Fano resonance are associated
with an asymmetrical flow of quasiparticles strongly related to the
valley-polarized currents of closed-system states, which has implications
for applications in ``valleytronic'' devices\citep{valleytronics}.
The ubiquity of this phenomenon in the systems we have studied suggests
that they could appear in future experiments, and provides a motivation
for further theoretical and experimental work.
\begin{acknowledgments}
This research was conducted with funding from the Department of Energy
Computer Science Graduate Fellowship program under Contract No. DE-FG02-97ER25308.
MFB and EJH were supported by the Department of Energy, office of
basic science (grant DE-FG02-08ER46513).
\end{acknowledgments}
\bibliographystyle{unsrt}

\begin{thebibliography}{10}

\bibitem{Geim-Nature}
A.~K. Geim and K.~S. Novoselov.
\newblock The rise of graphene.
\newblock {\em Nature Materials}, 6(3):183--191, 03 2007.

\bibitem{Experiment1}
Melinda~Y. Han, Barbaros \"Ozyilmaz, Yuanbo Zhang, and Philip Kim.
\newblock Energy band-gap engineering of graphene nanoribbons.
\newblock {\em Phys. Rev. Lett.}, 98:206805, May 2007.

\bibitem{Experiment2}
K.~S. Novoselov, Z.~Jiang, Y.~Zhang, S.~V. Morozov, H.~L. Stormer, U.~Zeitler,
  J.~C. Maan, G.~S. Boebinger, P.~Kim, and A.~K. Geim.
\newblock Room-temperature quantum hall effect in graphene.
\newblock {\em Science}, 315(5817):1379, 2007.

\bibitem{Experiment3}
Elena Stolyarova, Kwang~Taeg Rim, Sunmin Ryu, Janina Maultzsch, Philip Kim,
  Louis~E. Brus, Tony~F. Heinz, Mark~S. Hybertsen, and George~W. Flynn.
\newblock High-resolution scanning tunneling microscopy imaging of mesoscopic
  graphene sheets on an insulating surface.
\newblock {\em Proceedings of the National Academy of Sciences},
  104(22):9209--9212, 2007.

\bibitem{Experiment5}
G.~M. Rutter, J.~N. Crain, N.~P. Guisinger, T.~Li, P.~N. First, and J.~A.
  Stroscio.
\newblock Scattering and interference in epitaxial graphene.
\newblock {\em Science}, 317(5835):219--222, 2007.

\bibitem{Experiment4}
Yuanbo Zhang, Victor~W. Brar, Caglar Girit, Alex Zettl, and Michael~F. Crommie.
\newblock Origin of spatial charge inhomogeneity in graphene.
\newblock {\em Nat Phys}, 5(10):722--726, 10 2009.

\bibitem{Mario}
J~Berezovsky, M~F Borunda, E~J Heller, and R~M Westervelt.
\newblock Imaging coherent transport in graphene (part i): mapping universal
  conductance fluctuations.
\newblock {\em Nanotechnology}, 21(27):274013, 2010.

\bibitem{Mario2-not-Mario}
Jesse Berezovsky and Robert~M Westervelt.
\newblock Imaging coherent transport in graphene (part ii): probing weak
  localization.
\newblock {\em Nanotechnology}, 21(27):274014, 2010.

\bibitem{castro-neto}
A.~H. Castro~Neto, F.~Guinea, N.~M.~R. Peres, K.~S. Novoselov, and A.~K. Geim.
\newblock The electronic properties of graphene.
\newblock {\em Rev. Mod. Phys.}, 81(1):109--162, Jan 2009.

\bibitem{Wimmer-Spin-Currents-Rough-Nanoribbons}
Michael Wimmer, \ifmmode \dot{I}\else \.{I}\fi{}nan\ifmmode
  \mbox{\c{c}}\else~\c{c}\fi{} Adagideli, Sava\ifmmode
  \mbox{\c{s}}\else~\c{s}\fi{} Berber, David Tom\'anek, and Klaus Richter.
\newblock Spin currents in rough graphene nanoribbons: Universal fluctuations
  and spin injection.
\newblock {\em Phys. Rev. Lett.}, 100(17):177207, May 2008.

\bibitem{KaxirasNanoFlakes}
Wei~L. Wang, Sheng Meng, and Efthimios Kaxiras.
\newblock Graphene nanoflakes with large spin.
\newblock {\em Nano Letters}, 8(1):241--245, 2008.
\newblock PMID: 18052302.

\bibitem{Wimmer-Spin-Transport}
Michael Wimmer, Matthias Scheid, and Klaus Richter.
\newblock {Spin-polarized Quantum Transport in Mesoscopic Conductors:
  Computational Concepts and Physical Phenomena}.
\newblock {\em arXiv:0803.3705}, 2008.

\bibitem{WeiLi}
Wei~L. Wang, Oleg~V. Yazyev, Sheng Meng, and Efthimios Kaxiras.
\newblock Topological frustration in graphene nanoflakes: Magnetic order and
  spin logic devices.
\newblock {\em Phys. Rev. Lett.}, 102(15):157201, Apr 2009.

\bibitem{molecular-doping-graphene}
T.~O. Wehling, K.~S. Novoselov, S.~V. Morozov, E.~E. Vdovin, M.~I. Katsnelson,
  A.~K. Geim, and A.~I. Lichtenstein.
\newblock Molecular doping of graphene.
\newblock {\em Nano Letters}, 8(1):173--177, 2008.
\newblock PMID: 18085811.

\bibitem{imaging-edge-states}
Stephan Schnez, Johannes G\"uttinger, Magdalena Huefner, Christoph Stampfer,
  Klaus Ensslin, and Thomas Ihn.
\newblock Imaging localized states in graphene nanostructures.
\newblock {\em Phys. Rev. B}, 82:165445, Oct 2010.

\bibitem{defect-graphene}
T.~O. Wehling, A.~V. Balatsky, M.~I. Katsnelson, A.~I. Lichtenstein,
  K.~Scharnberg, and R.~Wiesendanger.
\newblock Local electronic signatures of impurity states in graphene.
\newblock {\em Phys. Rev. B}, 75:125425, Mar 2007.

\bibitem{defect-point-standing-wave}
L.~Simon, C.~Bena, F.~Vonau, D.~Aubel, H.~Nasrallah, M.~Habar, and J.~C.
  Peruchetti.
\newblock Symmetry of standing waves generated by a point defect in epitaxial
  graphene.
\newblock {\em The European Physical Journal B - Condensed Matter and Complex
  Systems}, 69:351--355, 2009.
\newblock 10.1140/epjb/e2009-00142-3.

\bibitem{defects-graphene-point-theory-predict}
H.~Amara, S.~Latil, V.~Meunier, Ph. Lambin, and J.-C. Charlier.
\newblock Scanning tunneling microscopy fingerprints of point defects in
  graphene: A theoretical prediction.
\newblock {\em Phys. Rev. B}, 76:115423, Sep 2007.

\bibitem{katsnelson-ripples}
M.I Katsnelson and A.K Geim.
\newblock Electron scattering on microscopic corrugations in graphene.
\newblock {\em Philosophical Transactions of the Royal Society A: Mathematical,
  Physical and Engineering Sciences}, 366(1863):195--204, 2008.

\bibitem{reconstructing-edges}
Pekka Koskinen, Sami Malola, and Hannu H\"akkinen.
\newblock Evidence for graphene edges beyond zigzag and armchair.
\newblock {\em Phys. Rev. B}, 80:073401, Aug 2009.

\bibitem{imaging-edges}
Jifa Tian, Helin Cao, Wei Wu, Qingkai Yu, and Yong~P. Chen.
\newblock Direct imaging of graphene edges: Atomic structure and electronic
  scattering.
\newblock {\em Nano Letters}, 11(9):3663--3668, 2011.

\bibitem{edge-types-effects}
Kyle~A. Ritter and Joseph~W. Lyding.
\newblock The influence of edge structure on the electronic properties of
  graphene quantum dots and nanoribbons.
\newblock {\em Nature Materials}, 8(3):235--242, 03 2009.

\bibitem{Mason-PRL}
Douglas~J. Mason, Mario~F. Borunda, and Eric~J. Heller.
\newblock A semiclassical interpretation of probability flux.
\newblock (unpublished) arXiv:1205.0291.

\bibitem{Mason-Husimi-Continuous}
Douglas~J. Mason, Mario~F. Borunda, and Eric~J. Heller.
\newblock Extending the concept of probability flux.
\newblock (unpublished) arXiv:1205.3708.

\bibitem{Mason-Husimi-Lattices}
Douglas~J. Mason, Mario Borunda, and Eric~J. Heller.
\newblock Husimi projections in lattices.
\newblock (in preparation) arXiv:1206.1013, 2012.

\bibitem{PhysRevLett.53.1515}
Eric~J. Heller.
\newblock Bound-state eigenfunctions of classically chaotic hamiltonian
  systems: Scars of periodic orbits.
\newblock {\em Phys. Rev. Lett.}, 53(16):1515--1518, Oct 1984.

\bibitem{Huang-PRL-Scars}
Liang Huang, Ying-Cheng Lai, David Ferry, Stephen Goodnick, and Richard Akis.
\newblock {Relativistic Quantum Scars}.
\newblock {\em Phys. Rev. Lett.}, 103(5):1--4, July 2009.

\bibitem{fano-original}
U.~Fano.
\newblock Effects of configuration interaction on intensities and phase shifts.
\newblock {\em Phys. Rev.}, 124(6):1866--1878, Dec 1961.

\bibitem{JJAP.36.3944}
David~K. Ferry, Jonathan~P. Bird, Richard Akis, David P.~Pivin Jr., Kevin~M.
  Connolly, Koji Ishibashi, Yoshinobu Aoyagi, Takuo Sugano, and Yuichi Ochiai.
\newblock Quantum transport in single and multiple quantum dots.
\newblock {\em Jpn. J. Appl. Phys.}, 36(Part 1, No. 6B):3944--3950, 1997.

\bibitem{fano-review}
Andrey~E. Miroshnichenko, Sergej Flach, and Yuri~S. Kivshar.
\newblock Fano resonances in nanoscale structures.
\newblock {\em Rev. Mod. Phys.}, 82:2257--2298, Aug 2010.

\bibitem{Re-ZagDFT}
Pekka Koskinen, Sami Malola, and Hannu H\"akkinen.
\newblock Self-passivating edge reconstructions of graphene.
\newblock {\em Phys. Rev. Lett.}, 101:115502, Sep 2008.

\bibitem{Re-ZagWimmer}
J.~A.~M. van Ostaay, A.~R. Akhmerov, C.~W.~J. Beenakker, and M.~Wimmer.
\newblock Dirac boundary condition at the reconstructed zigzag edge of
  graphene.
\newblock {\em Phys. Rev. B}, 84:195434, Nov 2011.

\bibitem{ashcroft-and-mermin}
N.W. Ashcroft and N.D. Mermin.
\newblock {\em Solid state physics}.
\newblock Holt-Saunders International Editions: Science : Physics. Holt,
  Rinehart and Winston, 1976.

\bibitem{BeenakkerBC}
A.~R. Akhmerov and C.~W.~J. Beenakker.
\newblock Boundary conditions for dirac fermions on a terminated honeycomb
  lattice.
\newblock {\em Phys. Rev. B}, 77(8):085423, Feb 2008.

\bibitem{Wimmer-Robusteness}
M.~Wimmer, A.~R. Akhmerov, and F.~Guinea.
\newblock Robustness of edge states in graphene quantum dots.
\newblock {\em Phys. Rev. B}, 82(4):045409, Jul 2010.

\bibitem{ferry-pointer-states}
D.~K. Ferry, R.~Akis, and J.~P. Bird.
\newblock Einselection in action: Decoherence and pointer states in open
  quantum dots.
\newblock {\em Phys. Rev. Lett.}, 93:026803, Jul 2004.

\bibitem{Munoz-Rojas2006}
F.~Mu\~{n}oz Rojas, D.~Jacob, J.~Fern\'{a}ndez-Rossier, and J.~Palacios.
\newblock {Coherent transport in graphene nanoconstrictions}.
\newblock {\em Physical Review B}, 74(19):1--8, 2006.

\bibitem{Fano-Experiment2}
J.~G\"ores, D.~Goldhaber-Gordon, S.~Heemeyer, M.~A. Kastner, Hadas Shtrikman,
  D.~Mahalu, and U.~Meirav.
\newblock Fano resonances in electronic transport through a single-electron
  transistor.
\newblock {\em Phys. Rev. B}, 62:2188--2194, Jul 2000.

\bibitem{Fano-Experiment}
Kensuke Kobayashi, Hisashi Aikawa, Akira Sano, Shingo Katsumoto, and Yasuhiro
  Iye.
\newblock Fano resonance in a quantum wire with a side-coupled quantum dot.
\newblock {\em Phys. Rev. B}, 70:035319, Jul 2004.

\bibitem{Fano-Experiment3}
L.~E. Calvet, J.~P. Snyder, and W.~Wernsdorfer.
\newblock Fano resonance in electron transport through single dopant atoms.
\newblock {\em Phys. Rev. B}, 83:205415, May 2011.

\bibitem{datta}
S.~Datta.
\newblock {\em Electronic Transport in Mesoscopic Systems}.
\newblock Cambridge University Press, Cambridge, 1997.

\bibitem{mesoscopicbook}
L.L. Sohn, L.P. Kouwenhoven, and G.~Sch{\"o}n.
\newblock {\em Mesoscopic electron transport}.
\newblock NATO ASI series: Applied sciences. Kluwer Academic Publishers, 1997.

\bibitem{ferrybook}
D.K. Ferry and S.M. Goodnick.
\newblock {\em Transport in nanostructures}.
\newblock Cambridge Studies in Semiconductor Physics and Microelectronic
  Engineering. Cambridge University Press, 1999.

\bibitem{transport-book2}
I.O. Kulik and R.~Ellialt{\i}o{\u{g}}lu.
\newblock {\em Quantum mesoscopic phenomena and mesoscopic devices in
  microelectronics}.
\newblock NATO science series: Mathematical and physical sciences. Kluwer
  Academic, 2000.

\bibitem{dots-transport-book}
J.P. Bird.
\newblock {\em Electron transport in quantum dots}.
\newblock Kluwer Academic Publishers, 2003.

\bibitem{datta2}
S.~Datta.
\newblock {\em Quantum transport: atom to transistor}.
\newblock Cambridge University Press, 2005.

\bibitem{Wimmer-Energy-Levels}
J\"urgen Wurm, Adam Rycerz, \ifmmode \dot{I}\else \.{I}\fi{}nan\ifmmode
  \mbox{\c{c}}\else~\c{c}\fi{} Adagideli, Michael Wimmer, Klaus Richter, and
  Harold~U. Baranger.
\newblock Symmetry classes in graphene quantum dots: Universal spectral
  statistics, weak localization, and conductance fluctuations.
\newblock {\em Phys. Rev. Lett.}, 102:056806, Feb 2009.

\bibitem{Huang-Trans-Scar}
Liang Huang, Ying-Cheng Lai, David~K Ferry, Richard Akis, and Stephen~M
  Goodnick.
\newblock Transmission and scarring in graphene quantum dots.
\newblock {\em Journal of Physics: Condensed Matter}, 21(34):344203, 2009.

\bibitem{Huang-Open-Dots}
D~K Ferry, L~Huang, R~Yang, Y-C Lai, and R~Akis.
\newblock Open quantum dots in graphene: Scaling relativistic pointer states.
\newblock {\em Journal of Physics: Conference Series}, 220(1):012015, 2010.

\bibitem{Huang-Cond-Fluc}
Liang Huang, Rui Yang, and Ying-Cheng Lai.
\newblock Geometry-dependent conductance oscillations in graphene quantum dots.
\newblock {\em EPL (Europhysics Letters)}, 94(5):58003, 2011.

\bibitem{mason}
Douglas~J. Mason, David Prendergast, Jeffrey~B. Neaton, and Eric~J. Heller.
\newblock Algorithm for efficient elastic transport calculations for arbitrary
  device geometries.
\newblock {\em Phys. Rev. B}, 84:155401, Oct 2011.

\bibitem{graphene-bond-current2}
Jie-Yun Yan, Ping Zhang, Bo~Sun, Hai-Zhou Lu, Zhigang Wang, Suqing Duan, and
  Xian-Geng Zhao.
\newblock Quantum blockade and loop current induced by a single lattice defect
  in graphene nanoribbons.
\newblock {\em Phys. Rev. B}, 79:115403, Mar 2009.

\bibitem{graphene-bond-current}
L.~P. Z{\^a}rbo and B.~K. Nikoli{\'c}.
\newblock Spatial distribution of local currents of massless dirac fermions in
  quantum transport through graphene nanoribbons.
\newblock {\em EPL (Europhysics Letters)}, 80(4):47001, 2007.

\bibitem{valleytronics}
A.~Rycerz, J.~Tworzydlo, and C.~W.~J. Beenakker.
\newblock Valley filter and valley valve in graphene.
\newblock {\em Nat Phys}, 3(3):172--175, 03 2007.

\end{thebibliography}

\end{document}